\documentclass[prl,twocolumn,epsf,psfig]{revtex4}
\usepackage{graphicx}

\begin{document}
\title {Collective modes as a probe of the equation of state for partially polarized Fermi gases}
\author{Theja N. De Silva$^{a,b}$,
Erich J. Mueller$^{a}$}
\affiliation{$^{a}$ LASSP, Cornell University, Ithaca, New York 14853, USA. \\
$^{b}$ Department of physics, University of Ruhuna, Matara, Sri
Lanka.}

\begin{abstract}
We calculate the  collective modes of a partially polarized Fermi
gas trapped in a spherically symmetric harmonic potential. We show
that the breathing mode frequency exhibits non-monotonic dependence
on polarization in the entire BCS-BEC crossover region. Moreover, we
find that the breathing mode can be used to distinguish between two
commonly used unitary gas equations of state.
\end{abstract}

\maketitle

Recent experimental studies of partially polarized Fermi gases
\cite{randy,ketterle} at ultra-cold temperatures near a Feshbach
resonance provide a unique window into the behavior
 of strongly interacting fermions.  Relying on long spin relaxation times,
 experimentalists polarize their two-component Fermi gasses, driving them from superfluid to normal.
Consistent with theoretical predictions
\cite{theja2,theja3,chevy,theory1,theory2}, they see the cloud phase
separate into concentric shells, however, the detailed structure of
the shells are not completely understood.  The Rice experiments
\cite{randy} have largely been analyzed in terms of a two-shell
model where an unpolarized superfluid core is surrounded by a
completely polarized normal shell.  The MIT experiments
\cite{ketterle} strongly suggest a three-shell structure with a
partially polarized normal shell between the other regions.  There
currently exists no reliable model for the equation of state of
these partially polarized gases.  Here we explore to what extent
collective modes can be used to experimentally probe this equation
of state.

We use a hydrodynamic approach to calculate the breathing modes of a
partially polarized Fermi gas at unitarity.  We work with the two
most commonly used equations of state \cite{theja2,theja3,chevy},
finding a significant ($\sim5\%$) difference between the oscillation
frequencies.  This sensitivity should be contrasted to the case of
unpolarized gases where the differences between the predictions of
mean field theories and quantum monte-carlo are much smaller
($\sim2\%$) \cite{theja1,iso,modesT}.  Ongoing experimental efforts
which are attempting to observe these smaller beyond mean-field
effects \cite{modesE1,modesE2}, could easily distinguish between the
two equations of state which we consider, and can help refine our
theories of resonant fermions.

We study both the polarization and interaction strength dependance
of the breathing mode frequency. We find that the breathing mode
frequency is a non-monotonic  function of polarization for all
interaction strengths.

We consider a zero temperature gas of fermionic atoms of mass $m$ in
two hyperfine states $|\sigma=\uparrow,\downarrow>$ confined by a
spherically symmetric harmonic potential $U(r)=(m\omega_0^2/2)r^2$,
where $\omega_0$ is the trapping frequency. In terms of the numbers
of atoms $N_\sigma$ in each hyperfine state, the polarization is
defined as $P=(N_\uparrow-N_\downarrow)/(N_\uparrow+N_\downarrow)
\geq 0 $. Assuming local equilibrium, the dynamics of this system
will be described by a continuity equation $\partial_t
\rho_\sigma+\nabla\cdot j_\sigma=0$ and an Euler equation
$m\partial_t {\bf v}_\sigma+(m/2)\nabla v_\sigma^2+\nabla U_{\rm
eff}=0$, where $\rho_\sigma, {\bf v}_\sigma,$ and ${\bf
j}_\sigma=\rho_\sigma {\bf v}_\sigma$ are the mass density,
velocity, and mass current of the atoms in state $\sigma$. In the
local density approximation $U_{\rm eff}=U(r)-\mu_\sigma$.
Linearizing the equations around equilibrium density,
$\rho_\sigma=\rho_\sigma^0+\delta \rho_\sigma$, and writing the
density fluctuations in terms of fluctuations in chemical potential,
$\delta\rho_\sigma=\sum_\nu(\partial\rho_\sigma/\partial\mu_\nu)
\delta\mu_\nu$, we have
\begin{equation}\label{hydro}
\sum_\nu \kappa_{\sigma \nu}\partial_t^2 \delta\mu_\nu= \nabla\cdot
\left[\frac{\rho_\sigma}{m}\nabla\delta\mu_\sigma\right].
\end{equation}
\noindent where we have introduced the compressibility matrix
$\kappa_{st}=\partial\rho_s/\partial\mu_t$. Explicitly considering
the case of harmonic trapping, the equilibrium local chemical
potentials for up and down atoms in the local density approximation
(LDA) are given by $\mu_\uparrow(r)=\mu_0+h-(m\omega_0^2/2)r^2$ and
$\mu_\downarrow(r)=\mu_0-h-(m\omega_0^2/2)r^2$. If we know the
equation of state, we can extract the local equilibrium densities
from these local chemical potentials. Previous studies have shown
that this LDA approach is an excellent approximation unless the
number of particles is small and the trap deviates significantly
from spherical.  Under such circumstances, surface tension between
the superfluid and normal regions must be considered \cite{theja3}.
Here we neglect such finite size effects. The parameters $\mu_0$ and
$h$ are determined from a constraint on the total number of atoms
$N$ and the polarization $P$.

We formally reduce Eq.(\ref{hydro}) to a dimensionless form by
defining,
$\rho_\sigma=m(m\mu)^{3/2}f_\sigma(\mu/\epsilon,h/\epsilon)$,
$\kappa_{\sigma\nu}=m^2(m\mu)^{1/2}g_{\sigma\nu}(\mu/\epsilon,h/\epsilon)$,
and $x=\sqrt{m\omega_0^2/2\mu_0}r$, where $f_\sigma$ and
$g_{\sigma\nu}$ are dimensionless functions. The average chemical
potential $\mu=(\mu_\uparrow+\mu_\downarrow)/2=\mu_0(1-x^2)$ and
$\epsilon=\hbar^2/ma_s^2$ with s-wave scattering length $a_s$ and
Eq. (\ref{hydro}) becomes

\begin{equation}\label{hydro2}
\sum_\nu g_{\sigma
\nu}(\frac{\mu}{\epsilon},\frac{h}{\epsilon})\frac{\partial_t^2
\delta\mu_\nu}{\omega_0}= \frac{\nabla_x}{2}\cdot\left[
(1-x^2)^{\frac{3}{2}}f_\sigma(\frac{\mu}{\epsilon},\frac{h}{\epsilon})\nabla_x\delta\mu_\sigma\right].
\end{equation}
We see that dimensionless oscillation frequencies $\omega/\omega_0$
only depend on $\mu_0/\epsilon$ and $h/\epsilon$, or equivalently
$\sqrt{2m\mu_0/\hbar^2}a_s$ and $P$.

We work with two different equations of state: a BCS mean-field
result, and a semi-empirical model introduced by Frederic Chevy
\cite{chevy}.  The latter equation of state is sufficiently simple
that we can write the solutions of Eq.~(\ref{hydro}) in terms of
hypergeometric functions.  The BCS mean field theory is sufficiently
complicated that we must solve Eq.~(\ref{hydro}) numerically.  By
using spherical symmetry we reduce Eq.~(\ref{hydro}) to a
differential equation for the radial variation of
$\delta\mu_\sigma$.  After discretizing space and fourier
transforming with respect to time, this radial equation has the form
of a matrix eigenvalue problem.  Using standard sparse matrix
techniques we extract the eigenvalues and eigenvectors.  We verify
that our results are independent of the discretization procedure.

Neglecting the possibility of a modulated order parameter
(FFLO)\cite{fflo}, the mean-field equation of state is found by
numerically solving the following equations \cite{theja2} for
 the density $\rho=\rho_\uparrow+\rho_\downarrow$, the density difference $\rho_d=\rho_\uparrow-\rho_\downarrow$,
 and the gap $\Delta$,
 \begin{eqnarray}\label{reggap}
\frac{-m}{2\pi\hbar^2a_s} &=&  \int_0^{\infty} \frac{d^3\vec{k}}{(2
\pi)^3}\left(\frac{1}{E_k}-\frac{1}{
\epsilon_k}\right)-\int^{k_{+}}_{k_{-}} \frac{d^3\vec{k}}{(
2\pi)^3}\frac{1}{E_k}\\
\rho &=& \int_0^{\infty} \frac{d^3\vec{k}}{(2
\pi)^3}(1-\frac{\epsilon_k-\mu}{E_k})+\int^{k_{+}}_{k_{-}}(\frac{\epsilon_k-\mu}{E_k})\\
\rho_d&=&\frac{1}{(2 \pi)^3}\frac{4 \pi}{3}(k_{+}^3-k_{-}^3).
\end{eqnarray}
In these equations,
$E_{k}(\vec{r})=(E_{k\uparrow}+E_{k\downarrow})/2$, with
$E_{k\sigma}=\xi_{\sigma} h+\sqrt{(\epsilon_k-\mu)^2+\Delta^2}$,
where $\epsilon_k=\hbar^2 k^2/2m$,
$\xi_\uparrow=1$,$\xi_\downarrow=-1$, $\mu=
(\mu_{\uparrow}+\mu_{\downarrow})/2$ and
$h=(\mu_{\uparrow}-\mu_{\downarrow})/2$. The momenta of the Fermi
surfaces are $k_{\pm}(\vec{r})=(\pm\sqrt{h^2-\Delta^2}+\mu)^{1/2}$.
In the normal state, $\Delta=0$, this approach reproduces the
equation of state of a noninteracting Fermi gas. At unitarity, this
theory predicts a three-shell structure, similar to that observed at
MIT \cite{ketterle}, however the intermediate partially polarized
normal shell is predicted to be extremely small, and the mean-field
theory effectively drops all interactions in this shell.  Figure 1
shows the results of this hydrodynamic calculation at unitarity
$a_s\to\infty$, while figure 2 shows the results at several
different values of $a_s$.

Our second approximate equation of state was introduced by Frederic
Chevy \cite{chevy}, Unlike the mean-field theory, this approach
assumes that only two phases exist at unitarity: a completely
unpolarized superfluid, and  a completely polarized normal state.
The forms of the equations of state of these two phases are known
exactly: in the normal phase, $\rho=B\mu_\uparrow^{3/2}$, while in
the superfluid phase, $\rho=A[
(\mu_\uparrow+\mu_\downarrow)/2]^{3/2}$. Elementary statistical
mechanics gives $B=(1/6\pi^2)(2m/\hbar^2)^{3/2}$, while dimensional
analysis requires $A=2 B \varsigma^{-3/2}$. Comparison with
experiments and numerical Monte-Carlo calculations give
$\varsigma\approx0.45$ \cite{mc}. In this model, the edge of the
superfluid corresponds to the edge of the minority cloud
($R_\downarrow$).  At this boundary the ratio of the chemical
potential is a universal number,
$\mu_\downarrow/\mu_\uparrow=2(B/A)^{2/5}-1=-\xi\approx-0.061$
\cite{chevy,theja3}.  The edge of the majority species cloud
($R_\uparrow$) occurs when $\mu_\uparrow=0$, allowing us to write
$R_\downarrow^2=(2/m\omega_0^2)(\mu_0-h(1-\xi)/(1+\xi))$ and
$R_\uparrow^2=(2/m\omega_0^2)(\mu_0+h)$.

In the superfluid phase all four elements of the compressibility
matrix are equal. This structure results from the constraint
$\rho_\uparrow=\rho_\downarrow$. In the fully polarized phase, the
only nonzero element of the compressibility matrix is
$\kappa_{\uparrow\uparrow}$. Thus in each region the compressibility
matrix has rank 1, and there is only a single dynamical variable at
each point in space. In the superfluid phase ($r<R_\downarrow$), we
formalize this observation by taking the dot product (from the left)
of the vector $(1,-1)$ and Eq.~(\ref{hydro}). The time derivative
term vanishes and we are left with
$\nabla\cdot\left[(\rho_\uparrow/m)\nabla
(\delta\mu_\uparrow-\delta\mu_\downarrow) \right]=0$. A sufficient
condition for this to be satisfied is
$\delta\mu_\uparrow(r)-\delta\mu_\downarrow(r)=\lambda$, where
$\lambda$ is independent of $r$ (for $r<R_\downarrow$). Similarly,
in the polarized phase ($r>R_\downarrow$), $\rho_\downarrow=0$ and
we get $\rho_\downarrow \nabla\cdot\left[(\rho_\downarrow/m)\nabla
(\delta\mu_\downarrow) \right]=0$.  Again, a sufficient condition is
 $\delta\mu_\downarrow(r>R_\downarrow)=s$, where $s$ is a constant.
 Motion of the boundary is captured by the time dynamics of $s$ and $\lambda$.

To match the solutions in the two regions we note that at the edge
of the minority cloud, the density is discontinuous. This implies
that the compressibility matrix has a delta-function singularity.
Including this singularity in Eq. (\ref{hydro}), one sees that the
quantity in the divergence, $f_\sigma=\nabla
\delta\mu_\sigma/\rho_\sigma$, is discontinuous. By integrating Eq.
(\ref{hydro}) in the neighborhood of $R_\downarrow$, one finds that
this discontinuity is due solely to the discontinuity in
$\rho_\sigma$, and that both $\delta\mu_\uparrow$ and its first
derivative are continuous.

Equating the two expressions for $\delta\mu_\downarrow$ at the
boundary, we have $\delta\mu_\uparrow(R_\downarrow)-s=\lambda$.
Introducing
$w(r)=\delta\mu_\uparrow(r)-\delta\mu_\uparrow(R_\downarrow)$, we
then have,
\begin{eqnarray}\label{inside}
\partial_t^2(w+s+\lambda/2)=\frac{1}{2\kappa_{\uparrow\uparrow}}\nabla\cdot\left[\frac{\rho_\uparrow}{m}\nabla w\right],&&r<R_\downarrow,\\
\label{outside}
\partial_t^2(w+s+\lambda)=\frac{1}{\kappa_{\uparrow\uparrow}}\nabla\cdot\left[\frac{\rho_\uparrow}{m}\nabla w\right],&&r>R_\downarrow.
\end{eqnarray}
 Subtracting
Eq. (\ref{inside}) from (\ref{outside}) and setting $r=R_\downarrow$
we get an ordinary differential equation for $\lambda$, which is
readily solved in terms of $w(r=R_\downarrow)$, allowing us to
eliminate $\lambda$ from Eq. (\ref{inside}) and (\ref{outside}).
Finding the compressibilities from the equation of states, we derive
a closed set of equations for $r>R_\downarrow$ and $r<R_\downarrow$
respectively.
\begin{eqnarray}\label{outside2}
\frac{\partial_t^2\delta\mu_\uparrow(r)}{\omega_0^2}
&=&\frac{\left(R_\uparrow^2-r^2\right)}{3}\nabla^2\delta\mu_\uparrow(r)-
r \partial_r \delta\mu_\uparrow(r)\\\nonumber \label{inside2}
\frac{\partial_t^2\delta\mu_\uparrow(r)}{\omega_0^2}
&=&\frac{\left(\delta^2-r^2\right)}{3}\nabla^2\delta\mu_\uparrow(r)-
r \partial_r \delta\mu_\uparrow(r)+X.
\end{eqnarray}
\begin{figure}
\includegraphics[width=\columnwidth]{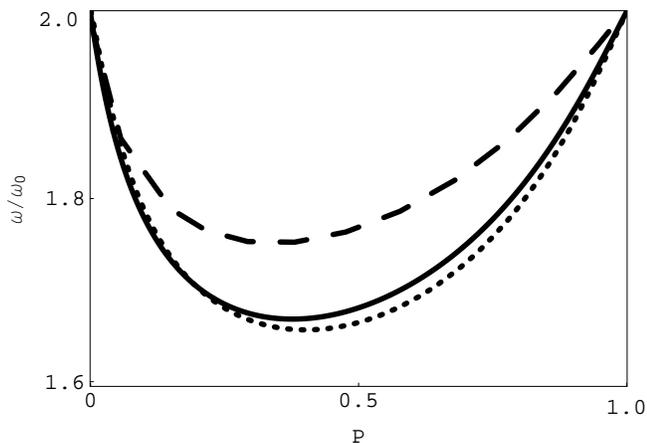}
\caption{Breathing mode frequencies of a spherical two component
Fermi gas as a function of polarization $P$ at unitarity. Solid
line: From matching the analytic solutions at the boundary between
superfluid and normal region in the two shell approximation. Dotted
line: From numerical solutions of Eq. (\ref{hydro}) using the
two-shell equation of state. Dashed line: Numerical solutions of Eq.
(\ref{hydro}) in the BCS approximation (which predicts a three shell
structure).}\label{unitary}
\end{figure}
where $\delta^2=R_\uparrow^2/(1+h/\mu_0)$ and
$3X/2=R^2\nabla^2\delta\mu_{\uparrow
+}(R_{\downarrow})-\Gamma^2\nabla^2\delta\mu_{\uparrow
-}(R_{\downarrow})$. Here $R^2=R_{\uparrow}^2-R_{\downarrow}^2$,
$\Gamma^2=\delta^2-R_{\downarrow}^2$, and $\delta\mu_{\uparrow
+/-}(R_{\downarrow})$ is the fluctuation of the majority species
chemical potential on the normal/superfluid side of the boundary.
Assuming $\delta\mu_\uparrow \propto \exp[i\omega t]$, and using a
suitable change of variables, each of the equations in Eq.
(\ref{outside2}) can be converted into hypergeometric equations and
the solutions are given in terms of Hypergeometric functions
$F(a,b;c,u)$ as \cite{hyper},
\begin{eqnarray}
\delta\mu_\uparrow&=&C_1
F(\alpha,\beta;\eta,1-r^2/R_\uparrow^2)r^lY_{l,m}(\theta,\phi)\\
\delta\mu_\uparrow&=&-\bar{X}+C_2
F(\alpha,\beta;\gamma,r^2/\delta^2)r^lY_{l,m}(\theta,\phi)
\end{eqnarray}
for $r>R_{\downarrow}$ and $r<R_{\downarrow}$ respectively. The
parameters $\alpha=(1/2)[2+l+\sqrt{4+3\epsilon+l^2+l}]$,
$\beta=(1/2)[2+l-\sqrt{4+3\epsilon+l^2+l}]$, $\gamma=(2l+3)/2$, and
$\eta=\alpha+\beta+1-\gamma$ are functions of magnetic quantum
number $l$ and the dimensionless mode frequencies
$\epsilon=\omega^2/\omega_0^2$. Here
$\bar{X}=(2/3)[R^2C_1G_1(R_\downarrow)-\Gamma^2C_2G_2(R_\downarrow)
]R_\downarrow^lY_{l,m}(\theta,\phi)$. Notice, we introduced two
arbitrary constants $C_1$ and $C_2$. The functions $G_1$ and $G_2$
are defined as,
$G_1(R_\downarrow)=(a/b)F(\alpha+2,\beta+2;\eta+2,1-R_\downarrow^2/R_\uparrow^2)
-(c/d)F(\alpha+1,\beta+1;\eta+1,1-R_\downarrow^2/R_\uparrow^2)$ and
$G_2(R_\downarrow)=(a/e)F(\alpha+2,\beta+2;\gamma+2,R_\downarrow^2/\delta^2)
+(c/(\gamma
\delta^2)F(\alpha+1,\beta+1;\gamma+1,R_\downarrow^2/\delta^2)$
\cite{note2}.

\begin{figure}
\includegraphics[width=\columnwidth]{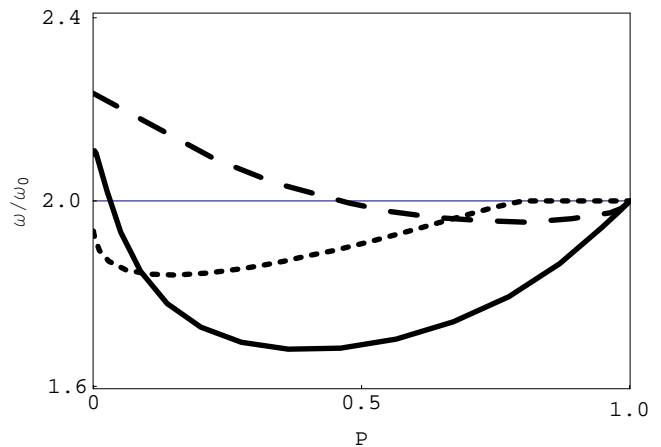}
\caption{Numerically calculated breathing mode frequencies in the
BCS approximation. Dotted line: BCS regime where theory predicts a
two/three shell structure; superfluid, partially polarized normal
and fully polarized normal shells. Solid line: Crossover-BEC regime
where theory predicts a two shell structure; superfluid and fully
polarized normal shells. Dashed line: Deep BEC regime where theory
predicts a two/three shell structure; superfluid, polarized
superfluid, and fully polarized normal shells. These three graphs
use the dimensionless parameter $\sqrt{2m\mu_0/\hbar^2}a_s$= -2.0,
+2.0, +0.2 respectively.}\label{bcs}
\end{figure}

We match the two solutions and their derivatives at the boundary to
get the frequencies $\omega$ for the breathing modes $(l=0)$ at
unitarity. The lowest energy breathing mode frequency is shown in
Fig. \ref{unitary} as a function of polarization. We compare these
results with those from the BCS equation of state.  We also
numerically solve Eq. (\ref{hydro}) using the two-shell equation of
state. The slight deviations of numerical and analytic results
obtained using the two shell equation of state are numerical
artifacts due to the smearing out of singularities in the
compressibilities at the boundary.

At both $P=0$ and $P=1$  the whole cloud is in a single phase with
an equation of state of the form $n\propto\mu^{3/2}$.  Consequently,
in both of these limits the cloud breaths with frequency
$\omega=2\omega_0$ \cite{iso}. For intermediate polarizations, the
mismatch of the speed of sound causes a drop in the oscillation
frequency.

Away from unitarity we must rely on the mean-field calculations, as
we know of no non-unitary analogy of Chevy's equation of state.   In
the BCS regime $(a_s<0)$, mean-field theory predicts a three-shell
structure at low polarizations: an unpolarized superfluid core
surrounded by partially polarized normal, and fully polarized normal
shells. At higher polarizations, the superfluid core is absent.
Since in the absence of a superfluid region the mean-field equation
of state reduces to that of non-interacting particles, one finds
that at these large polarizations the breathing mode frequency again
becomes $\omega=2\omega_0$ (see FIG.~\ref{bcs}). In the deep BEC
limit $( a_s>0,k_f a_s\ll1)$, the mean-field theory again predicts a
three-shell structure: an unpolarized superfluid core surrounded by
a partially polarized superfluid shell and a fully polarized normal
shell.  At high polarizations the inner core will be absent. In this
regime we do not see any dramatic signature of the disappearance of
the central superfluid core.

As seen in FIG. \ref{bcs}, the qualitative behavior of the
polarization dependence on breathing modes has the same
non-monotonic behavior in the entire BCS-BEC crossover region.

In ref. \cite{theja1}, we used a sum rule approach to find
collective mode frequencies for unpolarized Fermi gases. Although
the sum rule technique only provides upper bounds on these
frequencies, we found that the bound was very tight, and that method
produced excellent agreement with both experiments
\cite{modesE1,modesE2} and hydrodynamic theories \cite{iso,modesT}.
Repeating those calculations for the partially polarized gas, we
find that the sum rules provide a much weaker upper bound in the
present case.  For example, using the two-shell equation of state at
unitarity, the sum rule calculation finds no polarization
dependance.  It simply  bounds $\omega\leq2\omega_0$ for all P.

Although here we only consider spherically symmetric traps, we
believe that the qualitative behavior of the polarization dependence
of axial and radial breathing modes in an anisotropic traps will be
similar to what we found. In particular the frequencies of these
modes should be sensitive to the equation of state.  We speculate
that in highly asymmetric traps, such as those used at Rice, surface
tension effects may begin to play a role in the collective mode
frequencies, though we have not calculated these effects.

This work was supported by NSF grant PHY-0456261, and by the Alfred
P. Sloan Foundation.


\begin{references}
\bibitem{note2}The parameters $a$, $b$, $c$, $d$, and $e$ are defined as $a=4
R_\downarrow^2\alpha \beta(\alpha+1)(\beta+1)$,
$b=R_\uparrow^4\eta(\eta+1)$, $c=2\alpha \beta(2l+3)$,
$d=R_\uparrow^2\eta$, and $e=\delta^4\gamma(\gamma+1)$.
\end{references}
\end{document}